\documentclass[prb,twocolumn,superscriptaddress,showpacs]{revtex4}

\usepackage{graphicx}
\usepackage{amsfonts}

\newcommand{\mystate}[3]{\left| { {#1} \atop {#2}} #3\right>}
\newcommand{\ii}{\mathbf{\imath}}

\begin{document}

\title{Inter-shell exchange interaction in CdTe/ZnTe quantum dots:
 magneto-photoluminescence of X, X$^{2-}$ and XX$^-$}

\author{T. Kazimierczuk}
 \email{Tomasz.Kazimierczuk@fuw.edu.pl}
\author{T. Smole\'nski}
\affiliation{%
Institute of Experimental Physics, Faculty of Physics, University of
Warsaw,\\
Ho\.za 69, 00-681 Warsaw, Poland }%
\author{M. Goryca}
\affiliation{%
Institute of Experimental Physics, Faculty of Physics, University of
Warsaw,\\
Ho\.za 69, 00-681 Warsaw, Poland }%
\affiliation{%
Laboratoire National des Champs Magnétiques Intenses, \\
Grenoble High Magnetic Field Laboratory, \\
CNRS, 38042 Grenoble, France
}%

\author{\L{}. K\l{}opotowski}
\author{P. Wojnar}
\author{K. Fronc}
\affiliation{%
Institute of Physics, Polish Academy of Sciences,\\
Al. Lotnik\'ow 32/64, 02-688 Warsaw, Poland}%

\author{A. Golnik}
\author{M. Nawrocki}
\author{J.A. Gaj}
\thanks{deceased} 

\affiliation{%
Institute of Experimental Physics, Faculty of Physics, University of
Warsaw,\\
Ho\.za 69, 00-681 Warsaw, Poland }%

\author{P. Kossacki}
\affiliation{%
Institute of Experimental Physics, Faculty of Physics, University of
Warsaw,\\
Ho\.za 69, 00-681 Warsaw, Poland }%
\affiliation{%
Laboratoire National des Champs Magnétiques Intenses, \\
Grenoble High Magnetic Field Laboratory, \\
CNRS, 38042 Grenoble, France
}%

\date{\today}

\begin{abstract}
We present a comprehensive photoluminescence study of exchange interaction
in self-assembled CdTe/ZnTe quantum dots.  We exploit the presence of
multiple charge states in the photoluminescence spectra of single quantum
dots to analyze simultaneously fine structure of different excitonic
transitions, including recombination of
neutral exciton/biexciton, doubly charged negative exciton and negatively
charged biexciton. 
We demonstrate that the fine structure results from electron-hole exchange
interaction and that spin Hamiltonians with effective exchange constants
$\delta_i$ can provide a good description of each transition in magnetic
field for Faraday and Voigt field geometry. We determine and discuss values
of the effective exchange constants for a large statistics of quantum dots.
\end{abstract}


\maketitle

\section{Introduction}
Self-assembled quantum dots (QDs) are an ideal system when it comes to
study physics of closely confined carriers. One of interesting aspects of
such a system is the exchange interaction between the carriers. 
From a practical point of view, the electron-hole interaction in a single
QD is described using two characteristic energies: an isotropic
contribution responsible
for bright-dark exciton  splitting and anisotropic contribution related to
fine structure splitting of bright states of a neutral exciton
\cite{gammon-prl-1996}.
The studies of exchange interaction in QDs attracted wide attention
several years ago after the proposal of entangled photon generation in
biexciton-exciton (XX-X) cascade \cite{benson-prl-2000-qdentanglement}.
The interest was focused mainly on reducing anisotropic part of the e-h
exchange interaction, which hindered the entanglement between emitted
photons. The research effort finally led to successful demonstration of
fine structure control confirmed by observation of entanglement in XX-X
cascade\cite{stevenson-nature-2006,akopian-prl-2006}. 

In our present work we describe the exchange interaction in CdTe/ZnTe QDs.

Such dots are very convenient for spectroscopy as they give strong
photoluminescence (PL) in the visible range of the spectrum and exhibit
well resolved excitonic lines.
Another outstanding advantages of such dots is a feasibility of
incorporation a localized $5/2$ spins
by doping with manganese \cite{besombes-prl-2004-1mn}. A single Mn ion in
a QD was shown to exhibit long ($t_1 = 0.4$ms) spin memory
\cite{goryca-prl-2009-qd}, which makes it an interesting candidate for
quantum information storage.
Precise knowledge of the exchange interaction in such QDs is of the
essence due to its role in the optical orientation mechanism of the Mn spin
\cite{goryca-prb-2010}.
Our findings about electron-hole interaction in undoped self-assembled
CdTe/ZnTe QDs should be also valid for Mn-doped dots.

Most of the studies on e-h exchange were devoted to interactions between
$s$-shell carriers. Such an interaction is sufficient to describe a fine
structure of neutral exciton (X), charged excitons (X$^+$, X$^-$), and
neutral biexciton (XX). On the other hand, the fine structure of doubly
charged exciton (X$^{2-}$) and charged biexciton (XX$^-$) transitions is
determined by an exchange interaction between a $p$-shell electron and an
$s$-shell hole.
It was shown that this interaction can be successfully described using the
same approach as for the interaction between $s$-shell carriers, i.e., by
introducing respective
$p$-$s$ iso- and anisotropic exchange parameters as was demonstrated in
Refs.
\onlinecite{urbaszek-prl-2003,akimov-prb-2005,cade-prb-2006,ediger-prl-2007}.
However, most of the previous reports discuss effects related either to
X$^{2-}$ or XX$^-$ transitions. The simultaneous access to both complexes
allows the cross-comparison of all related fine structures, which is an
important test of the applicability limit of the used spin Hamiltonian
model. To our knowledge, such a comparison 
was reported only for highly symmetric QDs with negligible anisotropic
part of exchange interaction \cite{urbaszek-prl-2003}.

In this report, we present a comprehensive study of $s$-$s$ and $p$-$s$
electron-hole exchange interaction in CdTe/ZnTe QDs. The experiments
involved a large ($>200$) number of single dots and therefore the results
can be considered representative for the investigated system. 
In our study we focus particularly on transitions related to recombination
of X$^{2-}$ and XX$^-$ excitons. We show that fine structures of these
lines can be interpreted within a spin Hamiltonian featuring both iso- and
anisotropic term of exchange interaction. 
For simplicity, we assume that single-particle orbitals are not affected
by direct Coulomb interaction, which would be important for calculation of
absolute transition energies\cite{hawrylak-prb-1999}.
Moreover, we test the applicability of our model by introducing a magnetic
field either perpendicular (Faraday configuration) or parallel (Voigt
configuration) to the QD plane.
Finally, we compare the exchange parameters obtained independently from X,
X$^{2-}$, and XX$^-$ transitions to cross-check the consistency of our
description.

\section{Samples and experimental setup}

The samples were grown by molecular beam epitaxy (MBE) on GaAs substrate.
The sample structure contained four layers deposited during the growth: a
CdTe buffer (about 3$\mu$m), a ZnTe lower barrier (0.7 $\mu$m), a single
CdTe QD plane, and a ZnTe capping layer (50-100 nm). 
The QDs were formed following the original idea of Tinjod et. al.
\cite{tinjod-apl-2003}, in which a 2D CdTe layer is temporarily capped with
amorphous tellurium to induce the transition to dots.
A more detailed description of the sample growth can be found in Ref.
\onlinecite{wojnar-nanotechnology-2008}. 
Some of the samples were additionally post-processed by producing a gold
shadow-masks with 200nm apertures.
We did not find any significant differences in single dot properties
between different samples apart from the distribution of the QD emission
energies and the average charge state, which did not affect the results
presented in this work.

The sample was cooled to temperatures 1.5--10K. Spatial resolution defined
by the laser spot diameter was 0.5--2$\mu$m. The PL was excited
non-resonantly, using an Ar-ion laser (cw, 514nm), a Nd:YAG laser (cw,
532nm) or a frequency doubled Ti:Sapphire femtosecond laser (pulsed,
400nm). 
The choice of the excitation laser affected only intensities (both
absolute and relative) of the observed transitions and did not affect the
energy spectrum, thus the excitation details were not relevant to the
present work. The PL was analyzed using 0.3m - 0.5m spectrographs equipped
with CCD cameras and/or avalanche photodiode detectors. A $\lambda/2$
waveplate in a motorized mount followed by a linear polarizer allowed us to
perform repetitive automated measurements of polarization properties of QD
emission.

Most of the zero-field results were obtained using continuous-flow
cryostat with external microscope objective. The measurements exploiting
magnetic field were performed in cryostat equipped with a split-coil
superconducting magnet producing magnetic field up to 7T in Faraday or
Voigt configuration. Single photon correlations were performed using
Hanbury-Brown--Twiss detection scheme presented in detail in Ref.
\onlinecite{suffczynski-prb-2006}.
A complementary magneto-PL experiment using stronger magnetic field up to
28T was performed in Grenoble High Magnetic Field Laboratory. In the latter
case the measurement was not sensitive to the polarization of the PL
signal.

\section{Typical single QD photoluminescence pattern \label{sec:pattern}}

\begin{figure}
\includegraphics[width=85mm]{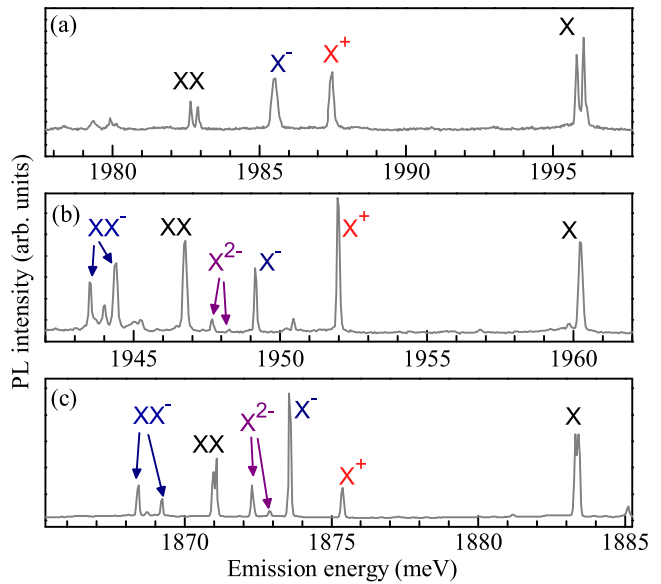}
\caption{Three examples of typical emission pattern of a CdTe/ZnTe QD.
Spectra were measured on different samples using non-resonant excitation
(532nm or 400nm) at low temperature ($<2$K). \label{fig:pl}}
\end{figure}

Studied dots were characterized by
relatively large span of emission energies from about 1800 meV (nearly 700
nm) to 2250 meV (about 550 nm). A typical single dot spectra are shown in
Fig. \ref{fig:pl}. Similar spectrum was observed by several groups
\cite{suffczynski-prb-2006,leger-prb-2007,lee-nanotechnology-2009}, however
only up to four strongest lines (X, X$^+$, X$^-$, XX) were recognized.
These lines tend to form a characteristic PL pattern with a single line
separated from the others. This single line in the high-energy side of the
PL spectrum is related to the neutral exciton, which can be confirmed e.g.
by the anisotropy measurement (see Section \ref{sec:anisotropy}). The next
two lines in the spectrum are related to the charged excitons (X$^+$,
X$^-$). The signs of their charge state were distinguished basing on the
charge tuning experiments on similar samples\cite{leger-prl-2006} and
observation of negative optical orientation under quasi-resonant
excitation\cite{kazimierczuk-prb-2009}. Such orientation was previously
found for negatively charged excitons in different material
systems\cite{neg1,neg2,neg3}. Finally, the last of the well established
four lines is related to the recombination of neutral biexciton. It is
evidenced by its polarization properties together with superlinear power
dependence.
Here we extend the description of the single dot spectrum by our
experimental results used to identify X$^{2-}$ and XX$^-$ transitions. These
transitions are closely related to $p$-$s$ electron-hole exchange
interaction but were not identified previously in CdTe/ZnTe system.

\begin{figure}
\includegraphics[width=85mm]{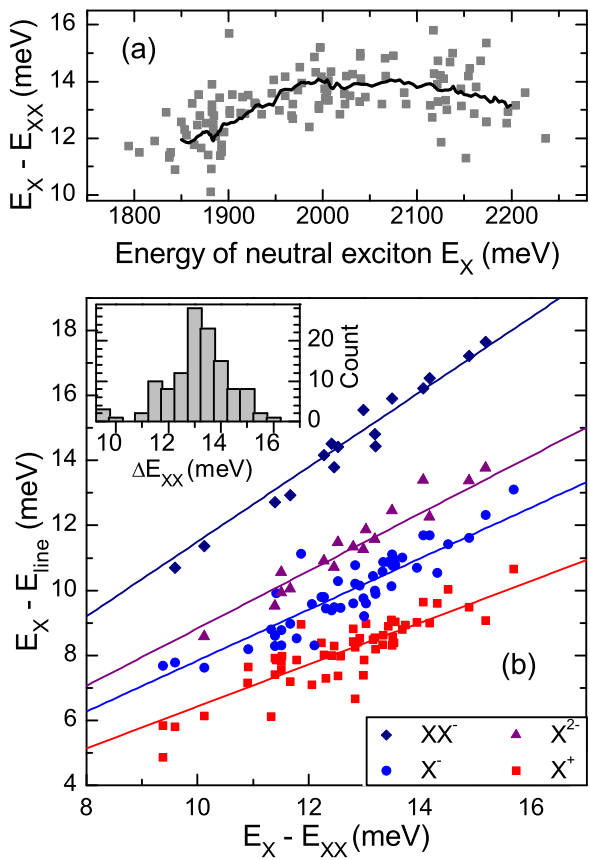}
\caption{(a) Correlation of XX relative energy and absolute emission
energy of X for different dots. A running average shown as a solid line is
drawn to guide the eye.
(b) Correlation of relative energies of various transitions for different
dots. Solid lines mark linear ($y=ax$) fits with proportionality constants
equal
$a=0.64$ for X$^+$, $a=0.78$ for X$^-$, $a=0.88$ for X$^{2-}$, and
$a=1.15$ for XX$^-$.
The inset presents distribution of XX relative energies.
\label{fig:binding}}
\end{figure}

We start with a discussion of relative energies (i.e. transition energies
with respect to transition energy of neutral exciton) of these transitions
in different dots. Such relative energies
vary between dots with large random scatter on top of systematic changes
with emission energy, as shown in Fig. \ref{fig:binding}(a).
On average, the values of relative energy of XX transition in our dots are
spread around 13.2 meV with standard deviation 1.3 meV (inset in Fig.
\ref{fig:binding}(b)).
We found that relative energies of various transitions for a single QD are
strongly correlated. Data collected in Fig. \ref{fig:binding}(b) shows a
clear linear dependence between relative energies of XX and
charged excitons (X$^+$, X$^-$, X$^{2-}$, and XX$^-$). This results in the
same transition sequence for all dots but scattered energetic spread of the
lines. Namely, the lines related to X$^{2-}$ transition are present between
X$^-$ and XX lines in the typical PL spectrum while the ones related to
XX$^-$ transition are below XX line.

\begin{figure}
\includegraphics[width=85mm]{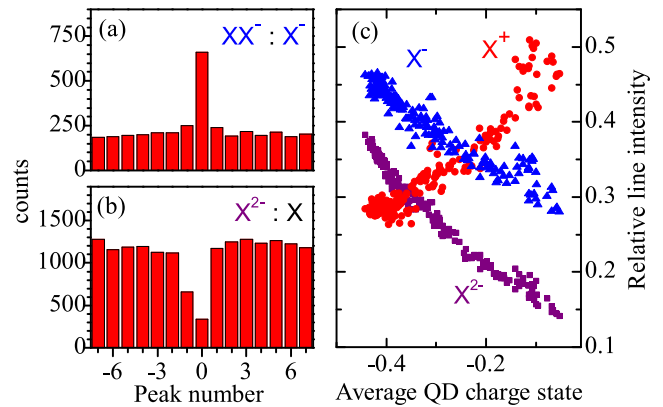}
\caption{
(a) Photon correlation histogram demonstrating XX$^-$--X$^-$ emission
cascade.
(b) Anti-bunching cross-correlation between X$^{2-}$ and X.
Negative peak number corresponds to detection of X$^{2-}$ after X.
Both correlation experiments were done with pulsed ps excitation of 400nm
with repetition 76MHz.
(c) PL intensities of charged excitons relatively to the intensity of the
neutral exciton. The estimate of the QD charge state is calculated as
$\sum_i q_iI_i/ \sum_i I_i$, where 
$I_i$ is the intensity of the transition related to charge state of $q_i$.
\label{fig:correlations}}
\end{figure}

A crucial point of our work is a correct identification of X$^{2-}$ and
XX$^-$ transitions. A conclusive argument for XX$^-$ identity was obtained
by means of single photon correlation measurement. 
Such a measurement performed with pulsed excitation gives relative
probability of the emission of two photons related to different transitions
in a single excitation event. In particular, strong correlation peak is
related to observation of two transitions from a single recombination
cascade (such as $\left|\mathrm{XX}\right> \to \left|\mathrm{X}\right> \to
\left|\emptyset\right>$).
A histogram evidencing cascade recombination of XX$^-$ and X$^-$ complexes
is shown in Fig. \ref{fig:correlations}(a). 

The photon cascade argument is not applicable to the case of X$^{2-}$
transition. Single photon measurements between previously identified
transitions and 
supposed X$^{2-}$ transition show clear
anti-bunching\cite{kazimierczuk-prb-2010}, which is expected for excitons
of different charge states. 
Also the asymmetric shape of the histogram is characteristic for the
correlation between charged and neutral excitons excited
non-resonantly\cite{suffczynski-prb-2006}.
The observed fine structure and magnetic field behavior (discussed further
in Sections \ref{sec:anisotropy} and \ref{sec:mag_x2min}) indicated doubly
charged exciton --- X$^{2-}$ or X$^{2+}$.
We determined the sign of the charge state by charge tuning experiment.
We exploited here the fact that the average charge state of studied QDs
depends on details of optical 
excitation, e.g., it can be modified by additional weak illumination with
high-energy light\cite{haas-jaszowiec-2009}.
In the experiment, we measured intensities of all excitonic transitions
while exciting a single QD simultaneously with two laser beams: 532nm and
(much weaker) 400nm.
By varying the intensity of 400nm laser beam we gradually changed the
average charge state estimated as $\sum_i q_iI_i/ \sum_i I_i$, where $I_i$
is an intensity of a transition related to charge state of $q_i$. The
results of such an experiment were presented in  Fig.
\ref{fig:correlations}(c). As expected X$^+$ and X$^-$ exhibit monotonic
(increasing and decreasing respectively) dependence on the average QD
charge state. Data for X$^{2-}$ follows the behavior of X$^{-}$, which
prove that these two complexes share the same sign of the charge state.

\section{Fine structure \label{sec:anisotropy}}

\begin{figure}
\includegraphics[width=85mm]{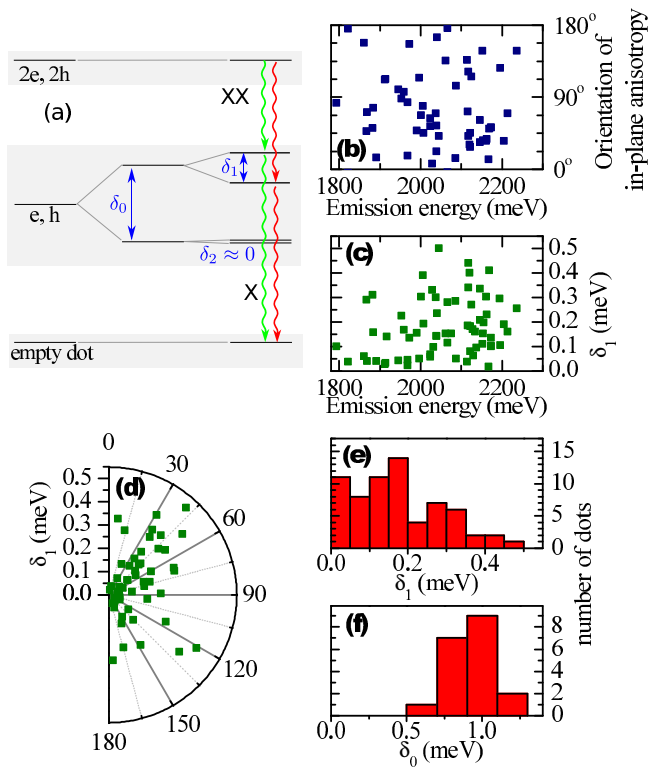}
\caption{(a) A schematic energy diagram illustrating the fine structure of
X and XX transitions.
(b-d) Correlations between various parameters showing random character of
in-plane anisotropy in CdTe/ZnTe dots: (b) orientation and (c) value of the
anisotropic splitting of X transition versus emission energy and (d) value
of anisotropic splitting and orientation of the in-plane anisotropy. (e)
Histogram of the anisotropic splitting $\delta_1$ of X transition. (f)
Histogram of bright--dark exciton splitting $\delta_0$.
\label{fig:anisostats}}
\end{figure}

Polarization resolved photoluminescence measurements represent an
important and very time-effective characterization tool. In general,
polarization dependence of the QD emission spectrum is related to in-plane
anisotropy of the confining potential. Possible physical origins of this
anisotropy
include non-cylindrical symmetry of the QD shape, in-plane strain or
electric field or the symmetry of the interfaces between barriers and the
QD\cite{kudelski-prb-2001}. The in-plane anisotropy was thoroughly studied
for many years because of its destructive role in the scheme of entangled
photon pair generation \cite{benson-prl-2000-qdentanglement}. It was shown
that a satisfactory description of the anisotropy-induced exciton splitting
can be achieved by introducing an anisotropy of exchange interaction
between electron and heavy hole. This interaction can be parametrized by
three quantities:  
\begin{eqnarray}
2 \langle\downarrow\Uparrow\mid H_\mathrm{exch} \mid \downarrow\Uparrow
\rangle = -2 
\langle\downarrow\Downarrow\mid H_\mathrm{exch} \mid \downarrow\Downarrow
\rangle & = & \delta_0 \\
2 \langle\downarrow\Uparrow\mid H_\mathrm{exch} \mid \uparrow\Downarrow
\rangle & = & \delta_1 \\
2 \langle\downarrow\Downarrow\mid H_\mathrm{exch} \mid \uparrow\Uparrow
\rangle & = & \delta_2
\end{eqnarray}
where $H_\mathrm{exch}$ is the effective exchange interaction and
$\downarrow$ and $\Downarrow$ 
represent z-component spin projection of the electron and the hole
respectively \cite{bayer-prb-2002}. 
These parameters are usually related to the fine structure of a neutral
exciton (Fig. \ref{fig:anisostats}(a)): splitting between dark and bright
branch ($\delta_0$), splitting between two bright configurations
($\delta_1$), and splitting between two dark excitons ($\delta_2$).
In the present work we neglect the presence of dark exciton splitting
(i.e., $\delta_2=0$).
This assumption is justified by very small values of $\delta_2$ (of order
of a few $\mu$eV \cite{poem-nature-2010}), never resolved in our
experiments.

We performed systematic measurements of PL polarization properties to
determine the influence
of the in-plane anisotropy on the typical emission pattern of single QDs
in our samples. 
In the experiment we recorded the PL spectra for a number of linear
polarization directions for each
studied dot. Such a procedure was necessary due to a large scatter of the
anisotropy axis between different QDs, usually observed in II-VI systems
\cite{kudelski-icps-2000}. The collected data allowed us to determine
actual principal axis of each QD and analyze the corresponding PL spectra.
For dots with small anisotropic splitting (smaller than our experimental
resolution), we determined the value of 
the splitting by fitting a gaussian profile to the whole collected dataset
as described in Ref. \onlinecite{kowalik-prb-2007}.

\begin{figure}
\includegraphics[width=85mm]{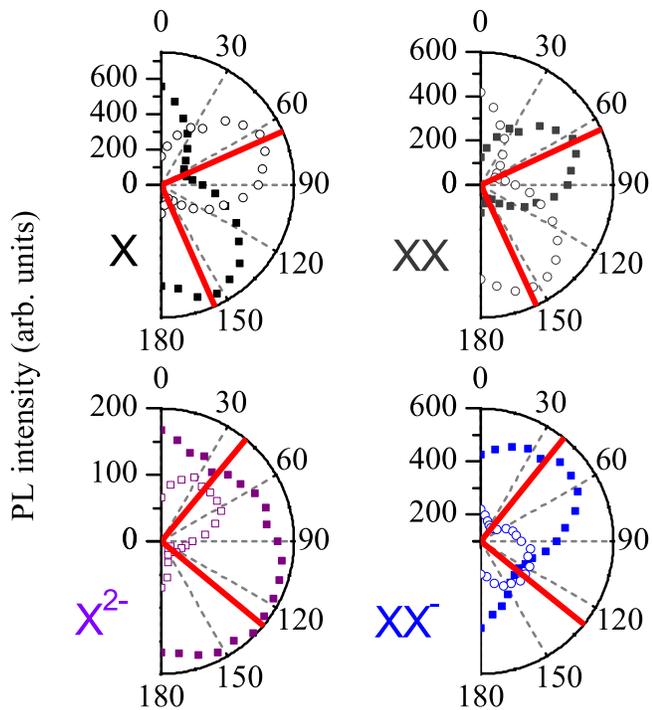}
\caption{Polar plots presenting orientation of anisotropy-related linear
polarization of various lines in the spectrum of a typical single QD.
Empty symbols are related to the higher-energy component of each
transition.
 \label{fig:aniso}}
\end{figure}

We start our discussion with analysis of the spectral lines related to the
recombination of the neutral excitonic complexes: X and XX. In the former
case, the zero-field emission consist of two closely spaced lines related
to spin configurations build from $\uparrow\Downarrow$ and
$\downarrow\Uparrow$ states\cite{bayer-prb-2002}. These two lines are
separated by the energy $\delta_1$ and are visible in two perpendicular
linear polarizations. The same underlying splitting of neutral exciton
state affects also XX transition which also features polarization-resolved
doublet split by $\delta_1$.
Different ordering of the components of XX and X transitions results from
different role of neutral exciton state in both cases: as final and initial
state of the transition respectively.
Figure \ref{fig:anisostats}(b-d) shows correlations of various
anisotropy-related parameters for a set of measured  dots. Coherently with
previous reports on the anisotropy in a similar system
\cite{kudelski-icps-2000}, we observe no preferential direction of the
in-plane anisotropy. Neither the splitting value nor the anisotropy
direction exhibit significant correlation with the transition energy or the
biexciton relative energy.

Single charged excitons (X$^+$, X$^-$) do not exhibit noticeable
fingerprints of in-plane anisotropy. This is expected since the
electron-hole exchange interaction influences neither
the initial (where two majority carriers are forming closed shell with
$S=0$) nor the final state (only one carrier left) of the transition.
Although some degree of linear polarization of charged exciton transitions
may arise due to valence band mixing \cite{leger-prb-2007}, we did not
concentrate on this effect in our study due to its negligible intensity in
the studied samples.

\begin{figure}
\includegraphics{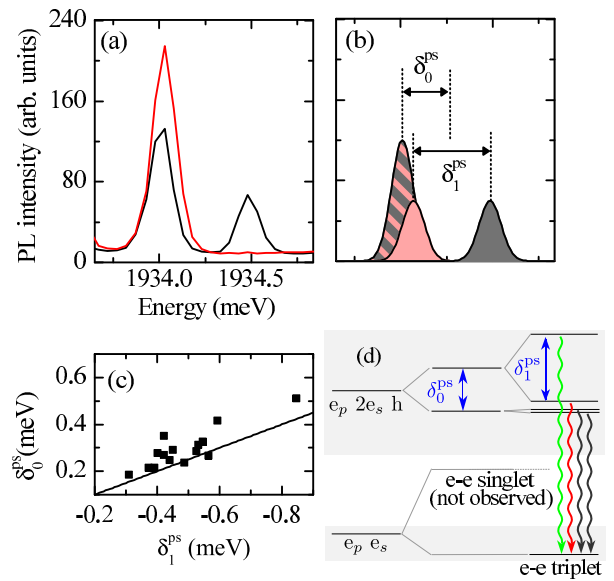}
\caption{(color online) (a) Typical photoluminescence signature of
X$^{2-}$ recombination in investigated dots. Black and red (gray) curves
correspond to two orthogonal linear polarizations. 
(b) Structure of the X$^{2-}$ emission spectrum. Curve fill denotes line
polarization (plain fill corresponds to full linear polarization, striped
pattern corresponds to unpolarized line). 
(c) A set of X$^{2-}$ exchange parameters for 15 different dots.
The line corresponds to relation $2\delta_0^{ps} = -\delta_1^{ps}$.
\label{fig:x2minus}
(d) A schematic illustration of the X$^{2-}$ fine structure in analogy to
the structure of the neutral exciton shown in Fig. \ref{fig:anisostats}(a).
The final state in the figure corresponds to the spin triplet
configuration.}
\end{figure}

The remaining two transitions --- X$^{2-}$ and XX$^-$ --- exhibit
more complex fine structure \cite{kazimierczuk-appa-2009}. Interestingly,
we observed that the orientation of the linear polarization of X$^{2-}$ and
XX$^-$ lines can be different than the orientation of the linear
polarization of X and XX lines (Fig. \ref{fig:aniso}). The mismatch between
these orientation is relatively small and varies from dot to dot.
Nevertheless, this difference clearly indicates a presence of additional
anisotropic interaction in X$^{2-}$ and XX$^-$ complexes. We identify it as
an exchange interaction between a hole and $p$-shell electron. 

In further considerations we tentatively assume that the in-plane
anisotropy significantly lifts the degeneracy between $p_x$ and $p_y$
orbitals. We also assume that the non-radiative relaxation of the excitonic
complex with an electron at the higher-energy orbital is much faster than
its optical recombination. In other words, we interpret 
the splitting patterns taking into account only single excited ($p$-shell)
level without any contribution from orbital angular momentum. The spin part
of the wavefunction is sufficient to describe all observed effects.

Recombination of X$^{2-}$ was previously studied in GaAs-based dots by
several groups with results qualitatively different in terms of light
polarization \cite{ediger-prl-2007,urbaszek-prl-2003,poem-prb-2007}. In our
samples, we found that the X$^{2-}$ transition consists mainly of two
emission lines separated by approximately $0.5$meV as shown in Fig.
\ref{fig:x2minus}(a). The higher-energy line
is significantly (5$\times$) weaker than the lower-energy one and is
completely linearly polarized, similarly to the case described in Ref.
\onlinecite{ediger-prl-2007}. The lower energy line exhibits a partial
linear polarization in the orthogonal direction. This splitting pattern
arises from the electron-hole exchange interaction in the initial state of
the transition. In the initial state, only two out of three electrons can
accommodate in the lowest \emph{s} orbital forming a closed shell. The
remaining electron on the ${p}$ shell interacts with the hole, similarly to
the interaction between carriers forming a neutral exciton. This analogy
allows us to understand the arising splitting pattern, however
one has to note differences in the exchange energies (denoted by
$\delta_0^{ps}$ and $\delta_1^{ps}$ 
for isotropic and anisotropic part respectively) and selection rules (each
state of X$^{2-}$ is bright). 

The final state of the considered transition consists of two electrons
which form either spin singlet or triplet configuration, typically
separated by several meV\cite{finley-prb-2001}. 
The optical transitions involve recombination of an electron and a hole of
opposite spin orientations, therefore the ``dark" configurations 
($\mystate{\uparrow}{\uparrow\downarrow}{\Uparrow}$,
$\mystate{\downarrow}{\uparrow\downarrow}{\Downarrow}$) recombine to the
triplet states 
($\mystate{\uparrow}{\uparrow}{}$, $\mystate{\downarrow}{\downarrow}{}$).
The ``bright" states can recombine either to singlet or triplet state.
Therefore, for the spin singlet final state we expect only a pair of lines
of equal intensity in the PL spectrum. Experimentally measured spectrum is
more complex, thus we suppose that the observed transitions are related to
the triplet configuration in the final state. 
Indeed, in such a case, one expect three emission lines (Fig.
\ref{fig:x2minus}(d)): unpolarized emission from the ``dark" state and two
linearly polarized lines from anisotropy-split ``bright" doublet.
One should note that ``bright"/``dark" labels are used here only in
analogy to the case of neutral exciton. Actual intensity of ``dark" exciton
recombination is two times stronger than ``bright" exciton recombination in
case of X$^{2-}$, as the oscillator strength of the ``bright" state
recombination is divided into spin singlet and triplet configurations of
the final state.

A typical PL spectrum of X$^{2-}$ and corresponding decomposition into
elementary transitions discussed above are presented in Figs.
\ref{fig:x2minus}(a-b). 
Experimental data indicate a close coincidence of energies of the
unpolarized line and lower component of the anisotropy-split doublet. Such a
coincidence can be expressed in terms of exchange parameters as a relation:
$2\left|\delta_0^{ps}\right|\approx\left|\delta_1^{ps}\right|$.
This property was observed for all our dots for which X$^{2-}$ transition
was seen.
Surprisingly, a similar coincidence was observed also for anisotropic
InAs/GaAs QD \cite{ediger-prl-2007}. No underlying reason for this
coincidence has been proposed so far.

In order to analyze quantitatively the relation between $\delta_0^{ps}$
and $\delta_1^{ps}$ we performed simultaneous fitting of spectra in both
polarizations. This procedure allowed us to separate isotropic and
anisotropic contributions to the exchange interaction. The results of the
fitting procedure are presented in Fig. \ref{fig:x2minus}(c). By averaging
the extracted values we established typical values of exchange parameters
as $\delta_0^{ps} = \left(0.29\pm 0.08\right)$ meV and  $\delta_1^{ps} =
\left(-0.49\pm 0.12\right)$ meV.

Negative sign of $\delta_1^{ps}$ is related to the fact, that the 
$\frac{1}{\sqrt{2}}\left(\mystate{\uparrow}{\uparrow\downarrow}{\Downarrow}
+ \mystate{\downarrow}{\uparrow\downarrow}{\Uparrow}\right)$ state 
has lower energy than
$\frac{1}{\sqrt{2}}\left(\mystate{\uparrow}{\uparrow\downarrow}{\Downarrow}
- \mystate{\downarrow}{\uparrow\downarrow}{\Uparrow}\right)$ state, while
in the case of the neutral exciton
$\frac{1}{\sqrt{2}}\left(\mystate{}{\uparrow}{\Downarrow} +
\mystate{}{\downarrow}{\Uparrow}\right)$ state has higher energy than 
$\frac{1}{\sqrt{2}}\left(\mystate{}{\uparrow}{\Downarrow} -
\mystate{}{\downarrow}{\Uparrow}\right)$ state. However, one has to take
into account the difference in the selection rules for X and X$^{2-}$.
Namely, the lines related to recombination of 
$\frac{1}{\sqrt{2}}\left(\mystate{\uparrow}{\uparrow\downarrow}{\Downarrow}
+ \mystate{\downarrow}{\uparrow\downarrow}{\Uparrow}\right)$ and
$\frac{1}{\sqrt{2}}\left(\mystate{}{\uparrow}{\Downarrow} +
\mystate{}{\downarrow}{\Uparrow}\right)$ states have two opposite linear
polarizations. This difference arises due to the fermionic nature of the
electrons\cite{poem-prb-2007} and is related to the sign of the $P_\pm$
matrices given in the Appendix \ref{appendixa}.  Therefore, in spite of
negative value of $\delta_1^{ps}$, the linearly polarized emission lines of
X and X$^{2-}$ exhibit the same order in the PL spectrum.

\begin{figure}
\includegraphics{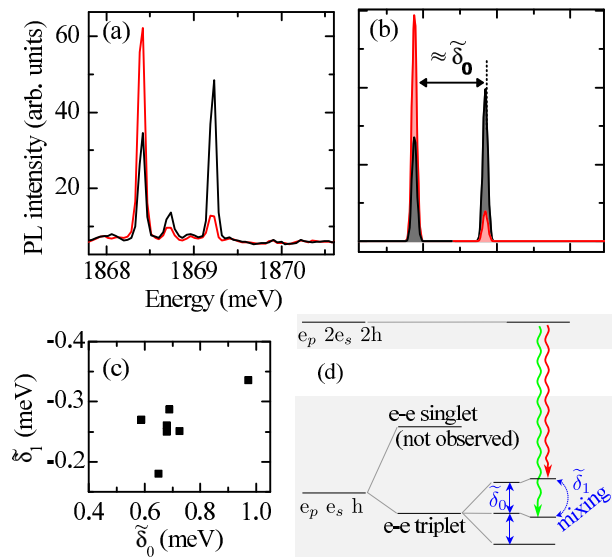}
\caption{(color online) (a) Typical PL signature of XX$^{-}$
recombination. Black and red (gray) curves correspond to two orthogonal
linear polarizations. 
(b) Simulated XX$^-$ emission spectrum. Curve fill denotes polarization  
(c) Correlation between two figures of merit for XX$^-$: 
$\widetilde{\delta_1}$ and $\widetilde{\delta_0}$. The presented data were
obtained by fitting field dependence as discussed in Section
\ref{sec:mag_xxmin}.
(d) Schematic energy diagram illustrating the origin of XX$^-$ fine
structure  \label{fig:xxminus}}
\end{figure}

We studied the role  of the \emph{p}-shell electron also in the fine
structure of the
XX$^-$ transition. The typical spectrum of XX$^-$ transition is presented
in Fig. \ref{fig:xxminus}(a). It consist of two lines. The intensity ratio
of these lines is close to 2:1 in favor of the lower energy line. Both
lines exhibit partial linear polarization. 
In this case the fine structure is determined mainly by the final state.
In the initial state (3 electrons + 2 holes) the electron-hole exchange
interaction is 
cancelled by paired hole spins. Similarly to the previously discussed
X$^{2-}$ transition,
spectroscopic signature of XX$^-$ transition fits to the case of triplet
configuration of electrons remaining in the final state (see Fig.
\ref{fig:xxminus}(d)).
Its degeneracy is lifted by the exchange interaction with remaining hole
forming three nearly-equidistant levels. In the simple picture of a
symmetric dot, the splittings should be equal to 
$\widetilde{\delta}_0 = \frac{1}{2}\left(\delta_0+\delta_0^{ps}\right)$ as
they result from the interaction between both $s$ and $p$ electron with the
hole. Only two out of three (four out of six including Krammers degeneracy)
configurations of the final state are optically active, giving rise to two
emission lines of XX$^-$ transition in the PL spectrum. The forbidden
configurations correspond to parallel spin orientations of all confined
carriers (
$\mystate{\uparrow}{\uparrow}{\Uparrow}$ and
$\mystate{\downarrow}{\downarrow}{\Downarrow}$).

The in-plane anisotropy manifests itself as mixing of the abovementioned
states by off-diagonal element proportional to 
$\widetilde{\delta}_1=\frac{1}{2}\left(\delta_1+\delta_1^{ps}\right)$. In
general, this addition should account for (possible) misorientation between
anisotropy of s and p shells, however in most cases the experimentally
determined mismatch between the two orientations is small.
The strength of this mixing can be evaluated by measuring the degree of
linear polarization of the corresponding PL lines. By diagonalization of a
Hamiltonian including both isotropic and anisotropic part of the exchange
interaction we found that the splitting between two components of XX$^-$
transition is given by 
$\sqrt{\widetilde{\delta}_0^2 + \frac{1}{2}\widetilde{\delta}_1^2}$.
Linear polarization degree ($P=\frac{I_\perp - I_{||}}{I_{\perp}+I_{||}}$)
of both lines is given by:
\begin{equation}
P = \frac{4 \beta}{1\pm 3\sqrt{1+2\beta^2}} \label{eq:linear_xxmin}
\end{equation}
where $\beta=\widetilde{\delta}_1 / \widetilde{\delta}_0$ and
sign `$+$' corresponds to the stronger of the two lines (i.e. line
corresponding to  
$\mystate{\uparrow}{\uparrow}{\Downarrow}$ and
$\mystate{\downarrow}{\downarrow}{\Uparrow}$
configurations in case of a symmetrical dot). 

The presented relations enable us to determine $\widetilde{\delta}_0$ and
$\widetilde{\delta}_1$ values separately. We found that in our dots the
average values of the exchange parameters were: 
$\widetilde{\delta}_0 = \left(0.75 \pm 0.16\right)$ meV and 
$\widetilde{\delta}_1 = \left(-0.28 \pm 0.08\right)$ meV.
Within a spin Hamiltonian picture used here, these values can be
independently  obtained by measuring separately electron-hole exchange
parameters for $s$- and $p$-shell electron. Using previously determined
$\delta_i$ and $\delta_i^{ps}$ values we found:
\begin{eqnarray}
\widetilde{\delta}_0^\mathrm{(calc)} & = & \frac{1}{2}\left(\delta_0 +
\delta_0^\mathrm{ps}\right) = \left(0.60 \pm 0.08\right) \mathrm{\ meV} \\
\widetilde{\delta}_1^\mathrm{(calc)} & = & \frac{1}{2}\left(\delta_1 +
\delta_1^\mathrm{ps}\right) = \left(-0.16 \pm 0.09\right) \mathrm{\ meV}
\end{eqnarray}
The values determined by combining the X and X$^{2-}$ fine structure
parameters are close to the values
obtained from analysis of XX$^-$ emission.
A small but not negligible difference between them gives a measure of
applicability of spin Hamiltonian approach to the fine structure of the
studied transitions.

\section{Magnetophotoluminescence}
Our study was completed by PL measurements in the magnetic field. 
The experiment was performed for two field configurations --- in-plane
(Voigt configuration) or along the growth axis (Faraday configuration).
In both cases, the influence of the magnetic field can be well described
by linear Zeeman term related to the spin and quadratic diamagnetic shift
related to the extension of the exciton wavefunction. In the present work
we were interested mainly in the influence of the magnetic field through
the Zeeman term on the fine structure of the excitonic states.

The magnitude of the magnetic interaction is governed by values of
electron and hole g-factors.
For simplicity, in further considerations we abstract from valence band
effects and introduce $\frac{1}{2}$ pseudospin for the $s$-shell hole with
effective anisotropic g-factor encapsulating e.g. heavy-light hole mixing.
In this convention the Zeeman splitting of the neutral exciton in Faraday
configuration is given by 
$\left(g_h^\mathbf{z}-g_e^\mathbf{z}\right)\mu_\mathrm{B}B$.
For the same purpose, we assume that $s$- and $p$-shell electrons are
characterized by the same g-factor. We also assume no contribution of the
orbital angular momentum of $p$-shell electron, possibly due to
anisotropy-related degeneracy lifting of two $p$ orbitals.

\subsection{Magnetophotoluminescence of X$^{2-}$ \label{sec:mag_x2min}}

\begin{figure}
\includegraphics{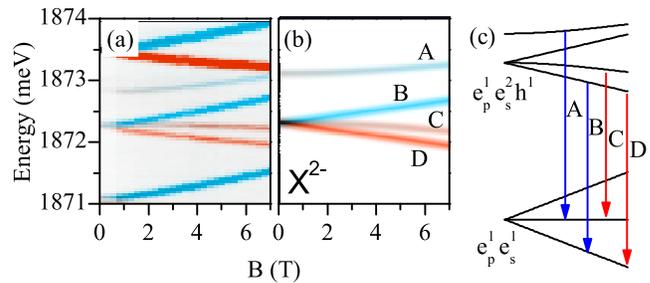}
\caption{(color online) (a) Experimental and (b) simulated PL spectrum of
X$^{2-}$ in magnetic field in Faraday configuration. Color saturation
denotes degree of circular polarization (blue --- $\sigma+$, red ---
$\sigma-$) and brightness denotes the PL intensity.
(c) Schematic of the corresponding energy diagram. 
\label{fig:x2min_faraday}}
\end{figure}

We start the discussion with the case of X$^{2-}$ transition in the
Faraday configuration. A typical 
data obtained in such an experiment is shown in Fig.
\ref{fig:x2min_faraday}(a).
The magnetic field splits the lower energy component of X$^{2-}$
transition into three lines (denoted B, C, D in Fig.
\ref{fig:x2min_faraday}(b)). The higher energy component (denoted A) does
not split. All four lines are naturally organised in two pairs, each
consisting of $\sigma+$ and $\sigma-$ polarized (fully or partially)
components of equal intensities. The stronger pair (B and D) is
characterized by splitting linear with field and it originates from the
lower energy component at $B=0$. Conversely, the other pair (A and C) is
already split at $B=0$ and the magnetic field induces only small increase
of the splitting. 

The observed behavior can be analyzed in analogy to neutral exciton,
invoked in the previous Section. In such a picture, the two pairs of lines
correspond to the two branches of the initial state: ``bright states'' and
``dark states''. The magnetic field acts on each branch independly. The
splitting of each branch is increasing according to 
$\sqrt{\delta^2 + \left(g\mu_\mathrm{B}B\right)^2}$ where $\delta$
corresponds to splitting at $B=0$ and $g$ is a respective excitonic Land\`e
factor. In such approach, the main difference between the two branches is
their zero-field splitting. Perfectly linear splitting of the ``dark
branch'' originates from negligible $\delta_2^{ps}$ value while substantial
zero-field splitting of the ``bright branch'' ($\delta_1^{ps}$) dominates
over the field-dependent contribution in the latter case. 

The presented analogy is instructive, however it is not perfect.
Particularly, it ignores the structure of the final state which is also
affected by the magnetic field. It is important especially for
recombination of ``dark states'', which according to selection rules lead
to $S=\pm 1$ branches of the final triplet state (Fig.
\ref{fig:x2min_faraday}(c)).
As a result, the optically observed splitting of a ``dark states'' is
governed by 
the same excitonic Land\`e factor as the ``bright states'' (i.e.,
$g_e^\mathbf{z}-g_h^\mathbf{z}$) and not the Land\`e factor of the real
dark neutral exciton (i.e., $g_e^\mathbf{z}+g_h^\mathbf{z}$).
This effect can be also seen as a result of the fact that the unpaired
electron from the initial state is not the same electron that is
recombining with the hole. 

\begin{table*}
\begin{tabular}{||c|c||}
\hline
\hline
Transition & Field dependence ($B || z$) from the spin Hamiltonian \\
\hline
\hline
X & $\frac{1}{2} \left( \delta_0\pm \sqrt{\delta_1^2+\left(
\left(g_e^\mathbf{z} - g_h^\mathbf{z}\right)  \mu_\mathrm{B}B \right)^2}
\right)$ \\
\hline 
X$^+$ and X$^-$ & $\pm\left(g_e^\mathbf{z} -
g_h^\mathbf{z}\right)\mu_\mathrm{B}B$ \\
\hline 
X$^{2-}$ & $\frac{1}{2} \left( \delta_0^{ps}\pm
\sqrt{\left(\delta_1^{ps}\right)^2+\left(\left(g_e^\mathbf{z} -
g_h^\mathbf{z}\right)\mu_\mathrm{B}B\right)^2}\right)$\\ 
 & $\frac{1}{2}\left(-\delta_0^{ps}\pm\left(g_e^\mathbf{z} -
g_h^\mathbf{z}\right)\mu_\mathrm{B}B \right)$ \\
\hline
XX$^-$ & $\frac{1}{2}\left(-\widetilde{\delta}_0 +
\sqrt{2\widetilde{\delta}_1^2+\left(\widetilde{\delta}_0 \pm
(g_e^\mathbf{z}-g_h^\mathbf{z})\mu_B B \right)^2} \right)$ \\
 & $\frac{1}{2}\left(-\widetilde{\delta}_0 -
\sqrt{2\widetilde{\delta}_1^2+\left(\widetilde{\delta}_0 \pm
(g_e^\mathbf{z}-g_h^\mathbf{z})\mu_B B \right)^2} \right)$ \\
\hline
\hline
\end{tabular}
\caption{Field dependence of transition energies in Faraday configuration
obtained from the spin Hamiltonian (neglecting diamagnetic shift).
\label{faraday_energies}}
\end{table*}

\begin{figure}
\includegraphics{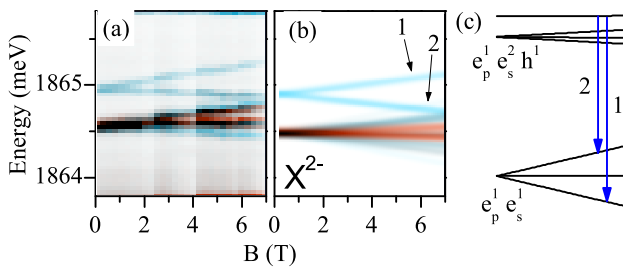}
\caption{(color online) (a) Experimental and (b) simulated PL spectrum of
X$^{2-}$ in magnetic field in Voigt configuration. Color saturation denotes
degree of linear polarization (red ---
$\vec{E}\parallel\vec{B}_\mathrm{ext}$, blue ---
$\vec{E}\perp\vec{B}_\mathrm{ext}$) and brightness denotes the PL
intensity. A mismatch between orientation of magnetic field and the QD
anisotropy was $16^\circ$.
(c) Schematic of the corresponding energy diagram. Transitions related to
characteristic splitting of higher-energy component were marked with
arrows.
 \label{fig:x2min_voigt}}
\end{figure}
PL measurements in Voigt configuration revealed qualitatively different
behavior of X$^{2-}$ transition (Fig. \ref{fig:x2min_voigt}(a)). This was
confirmed by repetition of the measurements for a number of dots with
different relative orientations of the in-plane anisotropy axis and the
magnetic field. In each case we observed splitting of the higher-energy
component into two lines and difficult to resolve multiple splitting of
lower-energy component. 

This spectral features are completely reproduced by the model based on
spin Hamiltonian given in Appendix \ref{appendixa} as shown in Fig.
\ref{fig:x2min_voigt}(b). 
The simulations confirm that the double splitting of the higher-energy
component does not depend on the relative orientation of in-plane
anisotropy. On the other hand, such a dependence was found in the spectral
lines split from the lower-energy component. Detailed analysis of the
involved energy levels allowed us to conclude that the two characteristic
lines split from the higher-energy component are transitions from the same
initial state to $S_\mathrm{x}=\pm 1$ states of the final triplet
configuration (Fig. \ref{fig:x2min_voigt}(c)). 
Thus, this splitting is a clear measure of a (doubled) in-plane g-factor
of an electron. Using this measure we found an average value of the
electron in-plane g-factor $g_e^\mathbf{x}=0.62$.

\subsection{Magnetophotoluminescence of XX$^{-}$ \label{sec:mag_xxmin}}

\begin{figure}
\includegraphics[width=85mm]{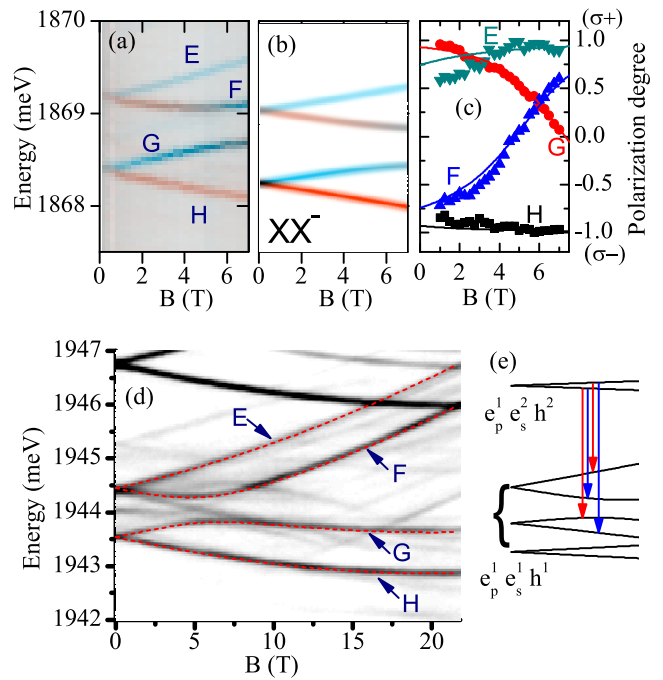}
\caption{(color online) (a) Magnetophotoluminescence of XX$^-$ transition
in Faraday configuration (field along the growth axis).
Blue and red represents $\sigma+$ and $\sigma-$ polarization respectively.
(b) Result of simulation using spin Hamiltonian given in Appendix
\ref{appendixa}.
(c) Degree of circular polarization for different components of XX$^-$ as
a function of magnetic field for the same QD. 
(d) Magnetophotoluminescence of XX$^-$ transition for a different dot
measured in a resistive magnet (no polarization resolution). Fitted model
is shown by dashed lines.
(e) Schematic of QD energy levels.  \label{fig:xxmin_faraday}}
\end{figure}

As it was shown in Section \ref{sec:anisotropy}, the zero field spectral
signature of XX$^-$ 
consists of two emission lines. Figure \ref{fig:xxmin_faraday}(a) presents
an evolution of this structure with external magnetic field in the Faraday
configuration.
Initially, for field up to a few tesla, both lines exhibit typical Zeeman
splitting into two circularly polarized components. 
However, the linearity of the Zeeman effect is perturbed
in the stronger field and an anticrossing between two inner spin-split
components is observed.
The anticrossing is accompanied by characteristic exchange of line
polarization, depicted in Fig. \ref{fig:xxmin_faraday}(c). 

The observed anticrossing is precisely reproduced by the model based on
the spin Hamiltonian (Fig. \ref{fig:xxmin_faraday}(b)). Similarly to the
fine structure at $B=0$, the field dependence is governed mainly by the
final states of the transitions. The anticrossing is due to the
off-diagonal anisotropic part of the e-h exchange interaction
($\widetilde{\delta}_1$). 
The model calculation reproduce also quantitatively the measured
polarization behavior (Fig. \ref{fig:xxmin_faraday}(c)).
It is worth to note that for none of the spectral lines the field
corresponding to complete linear polarization (and thus zero circular
polarization) coincides with the actual anticrossing point determined as a
point of minimum energy separation between the lines. 
Instead, at the anticrossing point both involved lines exhibit elliptical
polarization with the same contribution of $\sigma+$ polarization. Such an
effect is not related to the properties of the anticrossing states, but
rather to the difference in the values of transition matrix elements (see
Appendix \ref{appendixa}). 

The field dependence of XX$^-$ transition energies and particularly the
anticrossing strength is a direct measure of $\widetilde{\delta}_1$. We
verified that indeed the same pair of $\widetilde{\delta}_i$ values fits
transition energies and corresponding polarization degrees with and without
magnetic field. For example, the magnetic field measurements on a dot in
Fig. \ref{fig:xxmin_faraday}(a,c) allowed us to obtain
$\widetilde{\delta}_0=0.69$ meV and $\widetilde{\delta}_1=-0.29$ meV. By
substituting these values to Eq. \ref{eq:linear_xxmin} we predict degree of
zero field linear polarization for both PL lines of this particular dot as
$0.37$ and $-0.68$, while in the independent measurement we found them to
be equal to $0.36$ and $-0.69$.

Experiments in high magnetic field allowed us to compare the model
predictions with data measured in a wide range of applied field. Figure
\ref{fig:xxmin_faraday}(d) show example of the PL obtained in a scan up to
22T. The data clearly show the previously discussed anticrossing between
two bright transitions. In principle, another anticrossing is expected to
occur between 
the highest energy line and the optically inactive configuration of the
final state. It is expected to occur for field about 15T, but this value
strongly depends on the values of electron and hole g-factors, which are
difficult to access separately. During the experiment we found no evidence
of such an anticrossing. The lack of such an anticrossing is another (apart
from X$^{2-}$ fine structure) evidence of negligible value of
$\delta_2^{ps}$ exchange parameter.

\begin{figure}
\includegraphics{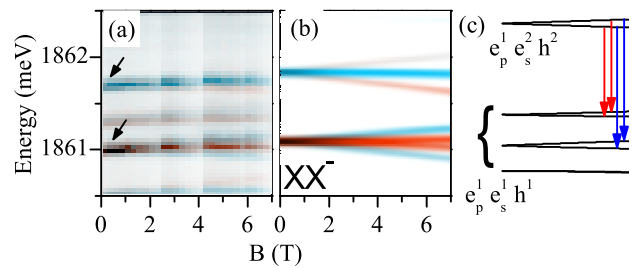}
\caption{(color online) (a) Experimental and (b) simulated PL spectrum of
XX$^-$ in magnetic field in Faraday configuration. Color saturation denotes
degree of linear polarization (red ---
$\vec{E}\parallel\vec{B}_\mathrm{ext}$, blue ---
$\vec{E}\perp\vec{B}_\mathrm{ext}$) and brightness denotes the PL
intensity. A mismatch between orientation of magnetic field and the QD
anisotropy was $16^\circ$. 
(c) Schematic of the corresponding energy diagram with the main
recombination channels. \label{fig:xxmin_voigt}}
\end{figure}
Finally, we studied the XX$^-$ transition in the Voigt geometry. Both the
experiment (Fig. \ref{fig:xxmin_voigt}(a)) and the model (Fig.
\ref{fig:xxmin_voigt}(b)) show only small variation of the PL spectrum of
XX$^-$. Interesting point in this configuration would be an observation of
the transition leading to optically forbidden branch of the final state, in
analogy to the dark neutral exciton, which is partially allowed by in-plane
magnetic field. The transition would produce emission lines at energy
approximately $\widetilde{\delta}_0$ above the higher energy component of
the XX$^-$ in the PL spectrum.  No such lines were found experimentally.
The reason is that in contrast to the case of neutral exciton, different
lines of XX$^-$ transition share the same initial state and a small
admixture of optically allowed states is not sufficient to successfully
compete with other radiative channels.

\section{Summary}

Basing on the set of different photoluminescence experiments we have
extracted a key parameters describing properties of excitonic complexes in
CdTe/ZnTe dots. The values of the parameters were analyzed statistically
over the large number of different dots. The averaged values of the main
parameters are summarized in Table \ref{tab:summary}.

\begin{table}
\begin{tabular}{|c|c|}
\hline
Parameter & Value \\
\hline
$E_{XX} - E_X$ & $\left(13.2 \pm 1.3\right)$ meV \\
$\delta_0$ & $\left(0.91 \pm 0.13 \right)$ meV \\
$\delta_1$ & $\left(0.18 \pm 0.11 \right)$ meV \\
$\delta_0^\mathrm{ps}$ & $\left(0.29 \pm 0.08 \right)$ meV \\
$\delta_1^\mathrm{ps}$ & $\left(-0.49 \pm 0.13\right)$ meV \\
\hline
\end{tabular}
\caption{A summary of parameters describing CdTe/ZnTe QD emission
spectrum. Symbols were explained in the text. Uncertainties of the listed
values are related to the spread between individual dots. They in each case
exceeded the experimental errors. \label{tab:summary}}
\end{table}

Despite large inhomogeneous broadening of QD emission in our samples, we
demonstrate 
an universality of a single-dot emission spectrum with characteristic
sequence of emission lines: X, X$^+$, X$^-$, X$^{2-}$ (two lines), XX, and
XX$^-$ (two lines). 
The fine structures of these transitions were successfully reproduced
using an extension of the model developed by Bayer et al
\cite{bayer-prb-2002}. The model involves exchange interaction between
$s$-shell electron and hole as well as between $s$-shell hole and $p$-shell
electron. 
We determined average values of parameters of these two interactions
separately by analysis of the fine structure of X and X$^{2-}$ transition
respectively.

Independently, from the analysis of XX$^-$ transition we have obtained
effective exchange parameters
$\widetilde{\delta_0} = \left(0.75 \pm 0.16\right)$ meV and
$\widetilde{\delta_1}=\left(-0.28 \pm 0.08\right)$ meV. We compare these
values with respective combination of $\delta_i$ and $\delta_i^{ps}$
parameters and find acceptable agreement between them. Such a comparison
is an important test of the consistency of the model based solely on spin
Hamiltonians. 

The second important result of our work is a verification of the model
calculations of excitonic transitions in the external magnetic field either
in Faraday or Voigt configuration. Our findings clearly demonstrate that
the measured transition energies as well as polarization selection rules
perfectly follow the theoretical predictions. This agreement firmly
supports applicability of the spin Hamiltonian model to the fine structure
of states featuring also $p$-shell electrons.

\begin{acknowledgments}
This work was supported by the Polish Ministry of Science
and Higher Education as research grants in years 2009-2011, by the
EuromagNetII, by the sixth Research Framework Programme of EU (Contract No.
MTKD-CT-2005-029671) and by the Foundation for Polish Science. One of us
(P.K.) was financially supported by the EU under FP7, Contract No. 221515
``MOCNA''.
\end{acknowledgments}

\appendix

\section{Hamiltonian operators \label{appendixa}}
Here we describe the Hamiltonians used for the calculation of energies of
excitonic states and transitions discussed in the manuscript.
For the sake of transparency of the calculations, we have included only
terms related to the observed effects. Namely, we included iso- and
anisotropic e-h exchange interaction (described by $\delta_0$ and 
$\delta_1$ for interaction with s-shell and 
$\delta_0^\mathrm{ps}$ and $\delta_1^\mathrm{ps}$ for interaction with
p-shell electron) and electron and hole g-factors (normal ---
$g_e^\mathbf{z}$ and $g_h^\mathbf{z}$; and in-plane ---
$g_e^{\mathbf{x},\mathbf{y}}$ and $g_h^{\mathbf{x},\mathbf{y}}$). We have
neglected exchange interaction between $\uparrow\Uparrow$ and
$\downarrow\Downarrow$ configurations ($\delta_2$ term), heavy-light hole
mixing, orbital effects related to the p-shell (zero-field degeneracy,
orbital angular momentum), anisotropy of e-e exchange interaction, and
configuration mixing of exciton complexes due to direct Coulomb
interaction\cite{hawrylak-prb-1999}. Furthermore, in the states containing
two electrons (initial state of X$^{2-}$ and final state of XX$^-$
transition) we have assumed dominant role of electron-electron interaction
and limited the analysis to the subspace corresponding to electron triplet
configuration. The orientation of the in-plane QD anisotropy is arbitrarily
chosen along the $x$ axis.

Eigenstates and their energies were obtained by analytical diagonalization
. The only exception was the final state of XX$^-$ transition with in-plane
magnetic field for which which the 6$\times$6 eigenproblem was solved
numerically.

Optical transitions were calculated using transition operators $P_+$ and
$P_-$ corresponding to
$\sigma_+$ and $\sigma_-$ polarization. We were not interested in the
absolute values of the oscillator strength and took $P_{+(-)} =
a_{s,\uparrow(\downarrow)} b_{s,\Downarrow(\Uparrow)}$ where $a$, $b$ are
annihilation operators for electron and hole respectively. Intensity
(relative) of PL line related to transition
$\left|i\right>\to\left|f\right>$ was calculated as 
$\left|\langle f\mid \alpha P_+ + \beta P_-\mid i\rangle\right|^2$ where
parameters $\alpha$ and $\beta$ are defined by the polarization used in
detection (e.g. $\alpha=1$, $\beta=0$ for $\sigma_+$ or
$\alpha=\beta=\frac{1}{\sqrt{2}}$ for horizontal linear polarization). In
such approach we did not analyze excitation dynamics nor the population
effect on the PL intensity.

Below we explicitly present all matrices related to X$^{2-}$ and XX$^-$
transitions. The base states are given using a notation
$\mystate{\mathrm{A}}{\mathrm{B}}{\mathrm{C}}$, where A is related to
$p$-shell electrons, B is related to $s$-shell electrons, and C is related
to $s$-shell holes.

\subsection{Matrices related to X$^{2-}$ transition}
Basis of the initial state:
\begin{displaymath}
\mystate{\uparrow}{\uparrow\downarrow}{\Uparrow}, \ 
\mystate{\downarrow}{\uparrow\downarrow}{\Uparrow}, \ 
\mystate{\uparrow}{\uparrow\downarrow}{\Downarrow}, \ 
\mystate{\downarrow}{\uparrow\downarrow}{\Downarrow}.
\end{displaymath}

Basis of the final state:
\begin{displaymath}
\mystate{\uparrow}{\uparrow}{}, \ 
\frac{1}{\sqrt{2}}\left(\mystate{\downarrow}{\uparrow}{}+
\mystate{\uparrow}{\downarrow}{}\right), \ 
\mystate{\downarrow}{\downarrow}{}.
\end{displaymath}

\begin{widetext}
Hamiltonian of the initial state  of X$^{2-}$ transition:
\begin{displaymath}
H_i = \frac{1}{2}\left(
\begin{array}{cccc}
-\delta_0^{ps} + \left(g_e^\mathbf{z} + g_h^\mathbf{z} \right) \mu_B B_z
  & g_e^\mathbf{x} \mu_B B_x - \ii g_e^\mathbf{y} \mu_B B_y
    & g_h^\mathbf{x} \mu_B B_x -\ii g_h^\mathbf{y} \mu_B B_y  
      & 0 \\
g_e^\mathbf{x} \mu_B B_x + \ii g_e^\mathbf{y} \mu_B B_y
  & \delta_0^{ps}+\left(-g_e^\mathbf{z} + g_h^\mathbf{z} \right) \mu_B B_z
    & \delta_1^{ps}
      & g_h^\mathbf{x} \mu_B B_x - \ii g_h^\mathbf{y} \mu_B B_y \\
g_h^\mathbf{x} \mu_B B_x + \ii g_h^\mathbf{y} \mu_B B_y
  & \delta_1^{ps}
    & \delta_0^{ps}+\left(g_e^\mathbf{z} - g_h^\mathbf{z} \right) \mu_B
B_z
      & g_e^\mathbf{x} \mu_B B_x - \ii g_e^\mathbf{y} \mu_B B_y\\
0
  & g_h^\mathbf{x} \mu_B B_x + \ii g_h^\mathbf{y} \mu_B B_y
    & g_e^\mathbf{x} \mu_B B_x + \ii g_e^\mathbf{y} \mu_B B_y
      & -\delta_0^{ps} +\left(-g_e^\mathbf{z} - g_h^\mathbf{z} \right)
\mu_B B_z  \\
\end{array} \right)
\end{displaymath}

Hamiltonian of the final state of X$^{2-}$ transition:
\begin{displaymath}
H_f = \left(
\begin{array}{ccc}
g_e^\mathbf{z} \mu_B B_z & \frac{1}{\sqrt{2}}g_e^\mathbf{x}\mu_B B_x -
\frac{\ii}{\sqrt{2}}g_e^\mathbf{y}\mu_B B_y  & 0 \\
\frac{1}{\sqrt{2}}g_e^\mathbf{x}\mu_B B_x + 
\frac{\ii}{\sqrt{2}}g_e^\mathbf{y}\mu_B B_y & 0 &
\frac{1}{\sqrt{2}}g_e^\mathbf{x}\mu_B B_x -
\frac{\ii}{\sqrt{2}}g_e^\mathbf{y}\mu_B B_y \\
0 & \frac{1}{\sqrt{2}} g_e^\mathbf{x}\mu_B B_x + 
\frac{\ii}{\sqrt{2}}g_e^\mathbf{y}\mu_B B_y & -g_e^\mathbf{z}\mu_B B_z 
\end{array} \right)
\end{displaymath}
\end{widetext}

Transition operators:
\begin{displaymath}
P_+ = \left(
\begin{array}{cccc}
-1 & 0 & 0 & 0 \\
0 & -\frac{1}{\sqrt{2}} & 0 & 0 \\
0 & 0 & 0 & 0
\end{array} \right)
\end{displaymath}
\begin{displaymath}
P_- = \left(
\begin{array}{cccc}
0 & 0 & 0 & 0 \\
0 & 0 & \frac{1}{\sqrt{2}} & 0 \\
0 & 0 & 0 & 1
\end{array} \right)
\end{displaymath}

\subsection{Matrices related to XX$^{-}$ transition}
Basis of the initial state:
\begin{displaymath}
\mystate{\uparrow}{\uparrow\downarrow}{\Uparrow\Downarrow}, \ 
\mystate{\downarrow}{\uparrow\downarrow}{\Uparrow\Downarrow}.
\end{displaymath}

Basis of the final state:
\begin{displaymath}
\begin{array}{l}
\mystate{\uparrow}{\uparrow}{\Uparrow}, \ 
\frac{1}{\sqrt{2}}\left(\mystate{\downarrow}{\uparrow}{\Uparrow}+
\mystate{\uparrow}{\downarrow}{\Uparrow}\right), \ 
\mystate{\downarrow}{\downarrow}{\Uparrow}, \\
\mystate{\uparrow}{\uparrow}{\Downarrow},\ 
\frac{1}{\sqrt{2}}\left(\mystate{\downarrow}{\uparrow}{\Downarrow}+
\mystate{\uparrow}{\downarrow}{\Downarrow}\right), \ 
\mystate{\downarrow}{\downarrow}{\Downarrow}.
\end{array}
\end{displaymath}

Hamiltonian of the initial state of XX$^{-}$ transition:
\begin{equation}
H_i = \frac{1}{2}\left(
\begin{array}{cc}
g_e^\mathbf{z} \mu_B B_z & g_e^\mathbf{x}\mu_B B_x - \ii
g_e^\mathbf{y}\mu_B B_y  \\
g_e^\mathbf{x}\mu_B B_x + \ii g_e^\mathbf{y}\mu_B B_y &
-g_e^\mathbf{z}\mu_B B_z 
\end{array} \right)
\label{eq:xxmin_hi}
\end{equation}

\begin{widetext}
Hamiltonian of the final state  of XX$^{-}$ transition:
\begin{equation}
H_f =
 \left(
\begin{array}{cccccc}
-\widetilde{\delta_0} + \xi_e^\mathbf{z} + \frac{1}{2}\xi_h^\mathbf{z}
  & \frac{1}{\sqrt{2}}\xi_e^\mathbf{x} -
\frac{\ii}{\sqrt{2}}\xi_e^\mathbf{y}
    & 0
      & \frac{1}{2}\xi_h^\mathbf{x} -\frac{\ii}{2}\xi_h^\mathbf{y}
        & 0
          & 0 \\
\frac{1}{\sqrt{2}}\xi_e^\mathbf{x} +\frac{\ii}{\sqrt{2}}\xi_e^\mathbf{y}
  & \frac{1}{2}\xi_h^\mathbf{z}
    & \frac{1}{\sqrt{2}}\xi_e^\mathbf{x} -
\frac{\ii}{\sqrt{2}}\xi_e^\mathbf{y}
      & \frac{1}{\sqrt{2}}\widetilde{\delta_1}
        & \frac{1}{2}\xi_h^\mathbf{x} -\frac{\ii}{2}\xi_h^\mathbf{y}
          & 0 \\
0 
  & \frac{1}{\sqrt{2}}\xi_e^\mathbf{x} +
\frac{\ii}{\sqrt{2}}\xi_e^\mathbf{y}
    & \widetilde{\delta_0}-\xi_e^\mathbf{z} + \frac{1}{2} \xi_h^\mathbf{z}
      & 0
        & \frac{1}{\sqrt{2}}\widetilde{\delta_1}
          & \frac{1}{2}\xi_h^\mathbf{x} - \frac{\ii}{2}\xi_e^\mathbf{y} \\
\frac{1}{2}\xi_h^\mathbf{x} + \frac{\ii}{2}\xi_h^\mathbf{y}
  & \frac{1}{\sqrt{2}}\widetilde{\delta_1}
    & 0
      & \widetilde{\delta_0} + \xi_e^\mathbf{z} -
\frac{1}{2}\xi_h^\mathbf{z}
        & \frac{1}{\sqrt{2}}\xi_e^\mathbf{x}
-\frac{\ii}{\sqrt{2}}\xi_e^\mathbf{y}
          & 0 \\

0
  & \frac{1}{2}\xi_h^\mathbf{x} + \frac{\ii}{2}\xi_h^\mathbf{y}
    & \frac{1}{\sqrt{2}}\widetilde{\delta_1}
      & \frac{1}{\sqrt{2}}\xi_e^\mathbf{x} +
\frac{\ii}{\sqrt{2}}\xi_e^\mathbf{y}
        & - \frac{1}{2}\xi_h^\mathbf{z}
          & \frac{1}{\sqrt{2}}\xi_e^\mathbf{x} -
\frac{\ii}{\sqrt{2}}\xi_e^\mathbf{y}\\

0
  & 0
    & \frac{1}{2}\xi_h^\mathbf{x} + \frac{\ii}{2}\xi_e^\mathbf{y}
      & 0
        & \frac{1}{\sqrt{2}}\xi_e^\mathbf{x} +
\frac{\ii}{\sqrt{2}}\xi_e^\mathbf{y}
          & -\widetilde{\delta_0} -\xi_e^\mathbf{z} -
\frac{1}{2}\xi_h^\mathbf{z}
\end{array} \right)
\label{eq:xxmin_hf}
\end{equation}
where $\xi_i^\mathbf{j}$ denotes $g_i^\mathbf{j}\mu_B B_j$ and
$\widetilde{\delta_i}$ denotes
$\frac{1}{2}\left(\delta_i+\delta_i^{ps}\right)$.
\end{widetext}

Transition operators:
\begin{displaymath}
P_- = \left(
\begin{array}{cccc}
0 & 0 \\
0 & 0 \\
0 & 0 \\
-1 & 0 \\
0 & -\frac{1}{\sqrt{2}} \\
0 & 0 \\
\end{array} \right),
P_+ = \left(
\begin{array}{cc}
0 & 0 \\
-\frac{1}{\sqrt{2}} & 0 \\
0 & -1 \\
0 & 0 \\
0 & 0 \\
0 & 0 \\
\end{array} \right).
\end{displaymath}

\section{Polarization of XX$^-$ lines in Faraday configuration
\label{appendixb}}
Below we give the analytical expressions describing the degree of circular
polarization for each spectral line of XX$^-$ transition in magnetic field
in Faraday configuration. 
The Hamiltonians used to obtain these expressions were presented in
Appendix \ref{appendixa} 
(Eq. \ref{eq:xxmin_hi} and \ref{eq:xxmin_hf}). 
$P_\mathrm{E}$, $P_\mathrm{F}$, $P_\mathrm{G}$, and $P_\mathrm{H}$ denote
circular polarization
$P=\frac{I_{\sigma+}-I_{\sigma-}}{I_{\sigma+}+I_{\sigma-}}$ for the lines
E, F, G, and H of XX$^-$ transition as defined in Fig.
\ref{fig:xxmin_faraday}.

\begin{eqnarray}
P_\mathrm{E} & = &
\frac{\left(\sqrt{2\beta^2+\left(\eta-1\right)^2}+\left(\eta-1\right)\right)^2-\beta^2}{\left(\sqrt{2\beta^2+\left(\eta-1\right)^2}+\left(\eta-1\right)\right)^2+\beta^2}
\\
P_\mathrm{F} & = &
\frac{\left(\sqrt{2\beta^2+\left(\eta+1\right)^2}+\left(\eta+1\right)\right)^2-4\beta^2}{\left(\sqrt{2\beta^2+\left(\eta+1\right)^2}+\left(\eta+1\right)\right)^2+4\beta^2}
\\
P_\mathrm{G} & = &
\frac{\left(\sqrt{2\beta^2+\left(\eta+1\right)^2}-\left(\eta+1\right)\right)^2-4\beta^2}{\left(\sqrt{2\beta^2+\left(\eta+1\right)^2}-\left(\eta+1\right)\right)^2+4\beta^2}
\\ \\
P_\mathrm{H} & = &
\frac{\left(\sqrt{2\beta^2+\left(\eta-1\right)^2}-\left(\eta-1\right)\right)^2-\beta^2}{\left(\sqrt{2\beta^2+\left(\eta-1\right)^2}-\left(\eta-1\right)\right)^2+\beta^2}
\\
\end{eqnarray}
where
\begin{eqnarray}
\beta & = & \widetilde{\delta_1}/\widetilde{\delta_0} \\
\eta & = & \left(g_e^\mathbf{z}-g_h^\mathbf{z}\right)\mu_B B /
\widetilde{\delta_0}
\end{eqnarray}

\end{document}